\documentstyle[12pt]{article}
\def\be {\begin{equation}}
\def\ee {\end{equation}}
\def\ba {\begin{eqnarray}}
\def\ea {\end{eqnarray}}

\def\bi {\begin{itemize}}
\def\ei {\end{itemize}}
\begin{document}
\def\bea{\begin{eqnarray}}
\def\eea{\end{eqnarray}}
\title{\bf {The holographic dark energy in non-flat Brans-Dicke cosmology}}
 \author{M.R. Setare  \footnote{E-mail: rezakord@ipm.ir}
  \\ {Department of Science,  Payame Noor University. Bijar, Iran}}
\date{\small{}}

\maketitle
\begin{abstract}
In this paper we study cosmological application of holographic dark
energy density in the Brans-Dicke framework. We employ
 the holographic model of dark energy to obtain the equation of state  for the holographic energy density
 in non-flat (closed) universe enclosed by
 the event horizon measured from the
 sphere of horizon named $L$. Our calculation show, taking $\Omega_{\Lambda}=0.73$ for
the present time, the lower bound of $w_{\rm \Lambda}$ is $-0.9$.
Therefore it is impossible to have $w_{\rm \Lambda}$ crossing $-1$.
This implies that one can not generate phantom-like equation of
state from a holographic dark energy model in non-flat universe in
the Brans-Dicke cosmology framework. In the other hand, we suggest a
correspondence between the holographic dark energy scenario in flat
universe and the phantom dark energy model in framework of
Brans-Dicke theory with potential.
 \end{abstract}

\newpage

\section{Introduction}
The accelerated expansion that based on recent astrophysical data
\cite{exp}, our universe is experiencing  is today's most
important problem of cosmology. Missing energy density - with
negative pressure - responsible for this expansion has been dubbed
Dark Energy (DE). Wide range of scenarios have been proposed to
explain this acceleration while most of them can not explain all
the features of universe or they have so many parameters that
makes them difficult to fit. The models which have been discussed
widely in literature are those which consider vacuum energy
(cosmological constant) \cite{cosmo} as DE, introduce fifth
elements and dub it quintessence \cite{quint} or scenarios named
phantom \cite{phant} with $w<-1$ , where $w$ is parameter of
state.

An approach to the problem of DE arises from holographic Principle
that states that the number of degrees of freedom related directly
to entropy scales with the enclosing area of the system. It was
shown by 'tHooft and Susskind \cite{hologram} that effective local
quantum field theories greatly overcount degrees of freedom
because the entropy scales extensively for an effective quantum
field theory in a box of size $L$ with UV cut-off $ \Lambda$. As
pointed out by \cite{myung}, attempting to solve this problem,
Cohen {\it et al.} showed \cite{cohen} that in quantum field
theory, short distance cut-off $\Lambda$ is related to long
distance cut-off $L$ due to the limit set by forming a black hole.
In other words the total energy of the system with size $L$ should
not exceed the mass of the same size black hole i.e. $L^3
\rho_{\Lambda}\leq LM_p^2$ where $\rho_{\Lambda}$ is the quantum
zero-point energy density caused by UV cutoff $\Lambda$ and $M_P$
denotes Planck mass ( $M_p^2=1/{8\pi G})$. The largest $L$ is
required to saturate this inequality. Then its holographic energy
density is given by $\rho_{\Lambda}= 3c^2M_p^2/8\pi L^2$ in which
$c$ is free dimensionless parameter and coefficient 3 is for
convenience.

 As an application of Holographic principle in cosmology,
 it was studied by \cite{KSM} that consequence of excluding those degrees of freedom of the system
 which will never be observed by that effective field
 theory gives rise to IR cut-off $L$ at the
 future event horizon. Thus in a universe dominated by DE, the
 future event horizon will tend to constant of the order $H^{-1}_0$, i.e. the present
 Hubble radius. The consequences of such a cut-off could be
 visible at the largest observable scales and particulary in the
 low CMB multipoles where we deal with discrete wave numbers. Considering the power spectrum in finite
 universe as a consequence of holographic constraint, with different boundary
 conditions, and fitting it with LSS, CMB and supernova data, a cosmic duality between dark energy equation of state
 and power spectrum is obtained that can describe the low $l$ features extremely
 well.

 Based on cosmological state of holographic principle, proposed by Fischler and
Susskind \cite{fischler}, the Holographic Model of Dark Energy
(HDE) has been proposed and studied widely in the
 literature \cite{miao,HDE}. In \cite{HG} using the type Ia
 supernova data, the model of HDE is constrained once
 when c is unity and another time when c is taken as free
 parameter. It is concluded that the HDE is consistent with recent observations, but future observations are needed to
 constrain this model more precisely. In another paper \cite{HL},
 the anthropic principle for HDE is discussed. It is found that,
 provided that the amplitude of fluctuation are variable the
 anthropic consideration favors the HDE over the cosmological
 constant.

 In HDE, in order to determine the proper and well-behaved system's IR cut-off, there are some
difficulties that must be studied carefully to get results adapted
with experiments that claim our universe has accelerated
expansion. For instance, in the model proposed by \cite{miao}, it
is discussed that considering particle horizon, $R_p$,
 \be
 R_p=a\int_0^t\frac{dt}{a}=a\int_0^a\frac{da}{Ha^2}
\ee
 as the IR cut-off, the HDE density reads to be
 \be
  \rho_{\Lambda}\propto a^{-2(1+\frac{1}{c})},
\ee
 that implies $w>-1/3$ which does not lead to accelerated
universe. Also it is shown in \cite{easther} that for the case of
closed
universe, it violates the holographic bound.\\

The problem of taking apparent horizon (Hubble horizon) - the
outermost surface defined by the null rays which instantaneously
are not expanding, $R_A=1/H$ - as the IR cut-off in the flat
universe, was discussed by Hsu \cite{Hsu}. According to Hsu's
argument, employing Friedman equation $\rho=3M^2_PH^2$ where
$\rho$ is the total energy density and taking $L=H^{-1}$ we will
find $\rho_m=3(1-c^2)M^2_PH^2$. Thus either $\rho_m$ and
$\rho_{\Lambda}$ behave as $H^2$. So the DE results pressureless,
since $\rho_{\Lambda}$ scales as like as matter energy density
$\rho_m$ with the scale factor $a$ as $a^{-3}$. Also, taking
apparent horizon as the IR cut-off may result the constant
parameter of state $w$, which is in contradiction with recent
observations implying variable $w$ \cite{varw}. In our
consideration for non-flat universe, because of the small value of
$\Omega_k$ we can consider our model as a system which departs
slightly from flat space. Consequently we respect the results of
flat universe so that we treat apparent horizon only as an
arbitrary distance and not as the system's IR cut-off.

 On the other hand taking the event horizon, $R_h$, where
 \be
  R_h= a\int_t^\infty \frac{dt}{a}=a\int_a^\infty\frac{da}{Ha^2}
 \ee
 to be the IR cut-off, gives the results compatible with observations for flat
 universe.

 It is fair to claim that simplicity and reasonability of HDE provides
 more reliable frame to investigate the problem of DE rather than other models
proposed in the literature\cite{cosmo,quint,phant}. For instance
the coincidence or "why now" problem is easily solved in some
models of HDE based on this fundamental assumption that matter and
holographic dark energy do not conserve separately, but the matter
energy density
decays into the holographic energy density \cite{interac}.\\
Some experimental data has implied that our universe is not a
perfectly flat universe and recent papers have favored the universe
with spatial curvature \cite{curve}. As a matter of fact, we want to
remark that although it is believed that our universe is flat, a
contribution to the Friedmann equation from spatial curvature is
still possible if the number of e-foldings is not very large
\cite{miao2}. Defining the appropriate distance, for the case of
non-flat universe has another story. Some aspects of the problem has
been discussed in \cite{miao2,guberina}. In this case, the event
horizon can not be considered as the system's IR cut-off, because
for instance, when the dark energy is dominated and $c=1$, where $c$
is a positive constant, $\Omega_\Lambda=1+ \Omega_k$, we find $\dot
R_h<0$, while we know that in this situation we must be in de Sitter
space with constant EoS. To solve this problem, another distance is
considered- radial size of the event horizon measured on the sphere
of the horizon, denoted by $L$- and the evolution of holographic
model of dark energy in non-flat universe is investigated.
\\
Because the holographic energy density belongs to a dynamical
cosmological constant, we need a dynamical frame to accommodate it
instead of general relativity. Therefore it is worthwhile to
investigate the holographic energy density in the framework of the
Brans-Dicke theory \cite{{gong},{mu}, {tor}}. Einstein's theory of
gravity may not describe gravity at very high energy. The simplest
alternative to general relativity is Brans-Dicke scalar-tensor
theory \cite{bd}. The recent interest in scalar-tensor theories of
gravity arises from inflationary cosmology, supergravity and
superstring theory.\\
In present paper, using the holographic model of dark energy in
non-flat universe, we obtain equation of state for holographic dark
energy density for a Brans-Dicke universe enveloped by  $L$ as the
system's IR cut-off. We show that Brans-Dicke theory  without scalar
potential correspond to quintessencial HDE, and it is impossible to
have phantom era in the theory under discussion. Then we explain how
the addition of scalar potential modifies the interpretation of such
model as HDE.
\section{Holographic Energy Density in Brans-Dicke Framework}

 The action of the Brans-Dicke theory in the canonical form is given
 by
\begin{equation}
S=\int d^{4}x\,\sqrt{g}\,\left[ -\frac{1}{8\omega }\,\phi ^{2}\,R+\frac{1}{2}%
\,g^{\mu \upsilon }\,\partial _{\mu }\phi \,\partial _{\nu }\phi
+L_{M}\right] .  \label{action*}
\end{equation}
where $\phi$ is the BD scalar. The non-minimal coupling term $\,\phi
^{2}\,R$ where $R$ is the Ricci scalar, replaces with the
Einstein-Hilbert term $\frac{1}{G_{N}}R$
in such a way that $G_{eff}^{-1}=\frac{2\pi }{\omega }\phi ^{2}$ where $%
G_{eff}$ is the effective gravitational constant as long as the
dynamical scalar field $\phi $ varies slowly \cite{arik}.
The gravitational field equations derived from the variation of the action (%
\ref{action*}) with respect to Robertson- Walker metric is
\begin{equation}
\frac{3}{4\omega }\,\phi ^{2}\,\left( \frac{\dot{a}^{2}}{a^{2}}+\frac{k}{%
a^{2}}\right) -\frac{1}{2}\,\dot{\phi}^{2}+%
\frac{3}{2\omega }\,\frac{\dot{a}}{a}\,\dot{\phi}\,\phi =\rho
\label{des}
\end{equation}%
\begin{equation}
\frac{-1}{4\omega }\phi ^{2}\left( 2\frac{\ddot{a}}{a}+\frac{\dot{a}^{2}}{%
a^{2}}+\frac{k}{a^{2}}\right) -\frac{1}{\omega }\,\frac{\dot{a}}{a}\,\dot{%
\phi}\,\phi -\frac{1}{2\omega }\,\ddot{\phi}\,\phi -\left( \frac{1}{2}+\frac{%
1}{2\omega }\right) \,\dot{\phi}^{2}=p \label{pres}
\end{equation}%
\begin{equation}
\ddot{\phi}+3\,\frac{\dot{a}}{a}\,\dot{\phi}-\frac{3}{2\omega }%
\left( \frac{\ddot{a}}{a}+\frac{\dot{a}^{2}}{a^{2}}+\frac{k}{a^{2}}\right) %
 \,\phi =0  \label{fi}
\end{equation}%
where $k$\ is the curvature parameter with $k=-1$, $0$, $1$\
corresponding to open, flat, closed universes respectively and
$a\left( t\right) $ is the scale factor of the universe. Here
$\rho = \rho_{\Lambda} + \rho_m$ and $p = p_{\Lambda} + p_m$.
Where $\rho_{\Lambda}$ and $\rho_m$ are respectively the
holographic energy density and energy density of matter. The
holographic energy density $\rho_{\Lambda}$ is chosen to be \be
\rho_{\Lambda}=\frac{3\phi^{2}}{4\omega L^2} \label{holo}\ee $L$
is defined as the following form:
\begin{equation}
 L=ar(t),
\end{equation}
here, $a$, is scale factor and $r(t)$ can be obtained from the
following relation
\begin{equation}\label{rdef}
\int_0^{r(t)}\frac{dr}{\sqrt{1-kr^2}}=\int_t^\infty
\frac{dt}{a}=\frac{R_h}{a},
\end{equation}
$R_h$ is event horizon. For closed universe we have (same
calculation is valid for open universe by transformation)
 \be
 r(t)=\frac{1}{\sqrt{k}} sin y.
 \ee
 Here $y\equiv \sqrt{k}R_h/a$.\\
 The critical energy density, $\rho_{cr}$, and the energy density of
curvature, $\rho_k$, are given by following relations
respectively:
\begin{eqnarray} \label{ro}
\rho_{cr}=\frac{3\phi^{2}H^2}{4\omega },\quad
\rho_k=\frac{3k\phi^{2}}{2a^2\omega}.
\end{eqnarray}
Now we define the dimensionless dark energy as \be
\Omega_{\Lambda}=\frac{\rho_{\Lambda}}{\rho_{cr}}=\frac{1}{L^2H^2}\label{omega}
\ee Using definition $\Omega_\Lambda$ and relation (\ref{ro}),
$\dot L$ gets: \be \label{ldot}
 \dot L= HL+ a \dot{r(t)}=\frac{1}{\sqrt{\Omega_\Lambda}}-cos y,
\end{equation}
Now let us consider the dark energy dominated universe. In this case
the dark energy evolves according to  its conservation law \be
\dot{\rho}_{\Lambda}+3H(\rho_{\Lambda}+P_{\Lambda})=0
\label{coneq}\ee Here we assume that $\phi/\phi_0=(a/a_0)^\alpha$,
where $\alpha=\beta \kappa$, $\beta=\sqrt{\frac{2}{2\omega+3}}$,
$\kappa=\sqrt{8\pi G}$. Substitute this relation into equation
(\ref{coneq}) and using Eq.(\ref{holo}), we obtain following
equation of state \be
P_{\Lambda}=\frac{-1}{3}(2\alpha+1+2\sqrt{\Omega_{\Lambda}}\cos
y)\rho_{\Lambda} \label{eqstat} \ee Therefore the index of the
equation of state is as \be
w_{\Lambda}=\frac{-1}{3}(2\alpha+1+2\sqrt{\Omega_{\Lambda}}\cos y)
\label{index} \ee  Because Brans-Dicke cosmology becomes standard
cosmology when $\omega\rightarrow \infty$, in this case
$\alpha\rightarrow 0$, according to this result, we see that the
holographic dark energy has the following index of the equation of
state in non-flat standard cosmology  \be
w_{\Lambda}=\frac{-1}{3}(1+2\sqrt{\Omega_{\Lambda}}\cos y)
\label{index1} \ee which is exactly the result of \cite{huang}. Then
$w_{\rm \Lambda}$ is bounded from below by \be w_{\rm
\Lambda}=-\frac{1}{3}(1+2\sqrt{\Omega_{\rm \Lambda}}) \label{bond}
\ee taking $\Omega_{\Lambda}=0.73$ for the present time, the lower
bound of $w_{\rm \Lambda}$ is $-0.9$. Therefore it is impossible to
have $w_{\rm \Lambda}$ crossing $-1$. This implies that one can not
generate phantom-like equation of state from a holographic dark
energy model in non-flat universe  in the Brans-Dicke cosmology
framework.
\section{Scalar-Tensor Theory of Gravity with Potential}
 The scalar-tensor theory of gravity with a non-zero potential in
 the Jordan frame is given by following action \cite{bnw70}
 \begin{equation}
S=\int d^4 x \sqrt{g}[{1\over 2} \Bigl (F(\Phi)~R -
g^{\mu\nu}\partial_{\mu}\Phi\partial_{\nu} \Phi \Bigr) - U(\Phi) +
L_M(g_{\mu\nu})]~, \label{L}
\end{equation}
where $L_m$ describes dustlike matter and $F(\Phi)>0$. For a flat
FRW universe, the background equations are then \cite{{per},{gan}}
\begin{eqnarray}
3FH^2 &=& \rho_m + {\dot \Phi^2\over 2} + U - 3H {\dot F}\ ,
\label{H2}\\
-2 F {\dot H} &=& \rho_m + \dot \Phi^2 + {\ddot F} - H {\dot F}\ .
\label{Hdot}
\end{eqnarray}
The equation for the dilaton field is as
\begin{equation}
\ddot \Phi +3H\dot\Phi +{dU\over d\Phi}- 3(\dot H + 2H^2)~{dF\over
d\Phi} =0~.
\end{equation}
Similar to the previous section we take
$F(\phi)=\frac{\phi^{2}}{4\omega}$, in this case the difference
between actions (\ref{L}) and (\ref{action*}) is the potential term
in (\ref{L}). From Eqs.(\ref{H2},\ref{Hdot}), one can obtain the
dark energy density and pressure respectively as following \be
\label{de}
\rho_{\Lambda}=\frac{\dot{\phi}^{2}}{2}+U-3H\dot{F}=\frac{\dot{\phi}^{2}}{2}+U-\frac{3}{2\omega}H\dot{\phi}\phi
\ee \be \label{dp} P_{\Lambda}=-F(2\dot{H}+3H^2)=
\frac{\dot{\phi}^{2}}{2}(1+\frac{1}{\omega})-U+\frac{H}{\omega}\dot{\phi}\phi+\frac{\ddot{\phi
}\phi}{2\omega}\ee From Eqs.(\ref{de},\ref{dp}) one can see that,
when $\omega\rightarrow \infty$, the energy density and pressure of
DE are corresponding quantities for quintessence model, it is not
strange, because the scalar-tensor theory of gravity with potential
is an extended quintessence model. Using Eqs.(\ref{de},\ref{dp}),
the equation of state for DE is as \be \label{eqes} w_{\rm
\Lambda}=\frac{P_{\Lambda}}{\rho_{\Lambda}}=\frac{\frac{\dot{\phi}^{2}}{2}(1+\frac{1}{\omega})-U+
\frac{H}{\omega}\dot{\phi}\phi+\frac{\ddot{\phi
}\phi}{2\omega}}{\frac{\dot{\phi}^{2}}{2}+U-\frac{3}{2\omega}H\dot{\phi}\phi}\ee
In this case, as have shown in \cite{{per},{gan}}, the scalar-tensor
gravity can allow crossing of the phantom divided barrier. Now we
show, how the addition of scalar potential modifies the
interpretation of such model as HDE.\\
In the flat universe, our choice for the holographic dark energy
density is \be \rho_{\Lambda}=\frac{3\phi^{2}}{4\omega R_{h}^{2}}
\label{holo1}\ee where $R_{h}$ is the event horizon \be
\label{event}R_h=a\int_{t}^{\infty}\frac{dt}{a} \ee the The
dimensionless dark energy is given by \be
\Omega_{\Lambda}=\frac{\rho_{\Lambda}}{\rho_{cr}}=\frac{1}{R_{h}^{2}H^2}\label{omega1}
\ee
 Now we reconstruct the
scalar potential (phantom potential)  in light of the holographic
dark energy. Again we assume that
$\frac{\phi}{\phi_{0}}=(\frac{a}{a_0})^{\alpha}$. According to the
forms of dark energy density  eq.(\ref{de}), one can easily derive
the scalar potential term as \be \label{cor}
U=\rho_{\Lambda}+\frac{\alpha}{2}(\frac{3}{\omega}-\alpha)H^{2}\phi^{2}
\ee
 If we establish the
correspondence between the holographic dark energy and scalar field,
then using Eqs.(\ref{holo1}, \ref{omega1}, \ref{cor}) we have \be
U(\phi)=[\frac{3}{2\omega}\Omega_{\Lambda}+\alpha(\frac{3}{\omega}-\alpha)]\frac{H^{2}\phi^{2}}{2}.
\ee Also, may be one expect to obtain phantom stage in the HDE model
in the framework of Brans-Dicke scalar-tensor theory with potential,
if one consider an infrared cutoff as a combination of the particle
and future horizon \cite{{odi1},{odi2}}.
\section{Conclusions}
In order to solve cosmological problems and because the lack of our
knowledge, for instance to determine what could be the best
candidate for DE to explain the accelerated expansion of universe,
the cosmologists try to approach to best results as precise as they
can by considering all the possibilities they have. It is of
interest to remark that in the literature, the different scenarios
of DE has never been studied via considering special similar
horizon, as in \cite{davies}, in the standard cosmology framework,
the apparent horizon, $1/H$, determines our universe while in
\cite{gong}, in the Brans-Dicke cosmology framework, the universe is
enclosed by event horizon, $R_h$. As we discussed in introduction,
for flat universe the convenient horizon looks to be $R_h$ while in
non-flat universe we define $L$ because of the problems that arise
if we consider $R_h$ or $R_p$ (these problems arise if we consider
them as the system's IR cut-off). In present paper, we studied $L$,
as the horizon measured from the sphere of the horizon as system's
IR cut-off. Then, by Considering the holographic energy density as a
dynamical cosmological constant, we have obtained the equation of
state for the holographic energy density in the Brans-Dicke
framework. Our calculation show, taking $\Omega_{\Lambda}=0.73$ for
the present time, the lower bound of $w_{\rm \Lambda}$ is $-0.9$.
Therefore it is impossible to have $w_{\rm \Lambda}$ crossing $-1$.
This implies that one can not generate phantom-like equation of
state from an interacting holographic dark energy model in non-flat
universe in the Brans-Dicke cosmology framework. In the other hand
according to the Refs. \cite{{per},{gan}}, one can explicitly
demonstrate that scalar-tensor theories of gravity can predict
crossing of the phantom divided barrier. But it has been known that
one has to consider  scalar-tensor gravity with a non-zero potential
obtain such crossing. Also in \cite{{odi1}, {odi2}} the generalized
holographic dark energy model was developed  that among these models
there are examples which may cross phantom boundary as well as
effective quintessence examples.\\
Similar to the models of \cite{{odi1}, {odi2}} may be one expect to
obtain phantom stage in the HDE model in the framework of
Brans-Dicke scalar-tensor theory, if one consider an infrared cutoff
as a combination of the particle and future horizon. In the other
term with the choice $L$ as the combination of future and particle
horizon, the HDE can be described by a phantom scalar field in
Brans-Dicke theory.
\\
As it was mentioned in introduction, $c$ is a positive constant in
holographic model of dark energy, and($c\geq1$). However, if $c<1$,
the holographic dark energy model also will behave like a phantom
model of DE the amazing feature of which is that the equation of
state of dark energy component $w_{\rm \Lambda}$ crosses $-1$.
Hence, we see, the determining of the value of $c$ is a key point to
the feature of the holographic dark energy and the ultimate fate of
the universe as well. However, in the recent fit studies, different
groups gave different values to $c$. A direct fit of the present
available SNe Ia data with this holographic model indicates that the
best fit result is $c=0.21$ \cite{HG}. Recently, by calculating the
average equation of state of the dark energy and the angular scale
of the acoustic oscillation from the BOOMERANG and WMAP data on the
CMB to constrain the holographic dark energy model, the authors show
that the reasonable result is $c\sim 0.7$ \cite{cmb1}. In the other
hand, in the study of the constraints on the dark energy from the
holographic connection to the small $l$ CMB suppression, an opposite
result is derived, i.e. it implies the best fit result is $c=2.1$
\cite{cmb3}.\\
Finally I must mention that in recent work \cite{nset} I  have
associated the interacting holographic dark energy in non-flat
universe with a phantom scalar field. I have shown that the
holographic dark energy with $c< 1$ can be described
 by the phantom in a certain way. A correspondence
between the holographic dark energy and phantom has been
established, and the potential of the holographic phantom and the
dynamics of the field have been reconstructed. In this regard in the
section $3$ of this paper we have suggested  a correspondence
between the holographic dark energy scenario in flat universe and
the phantom dark energy model in framework of Brans-Dicke theory
with potential.

\end{document}